\documentclass[11pt]{article}
\usepackage{amsmath}
\usepackage{amssymb}
\usepackage{epsfig}
\begin{document}
\title{{\bf{\Large Gravitational anomalies and one dimensional behaviour of black holes}}}
\author{ 
 {\bf {\normalsize Bibhas Ranjan Majhi}$
$\thanks{E-mail: bibhas.majhi@iitg.ernet.in}}\\ 
{\normalsize Department of Physics, Indian Institute of Technology Guwahati,}
\\{\normalsize Guwahati 781039, Assam, India}
\\[0.3cm]
}
\maketitle

\begin{abstract}
 It has been pointed out by Bekenstein and Mayo that the behavior of the Black hole's entropy or information flow is similar to that through one-dimensional channel. Here I analyse the same issue with the use of gravitational anomalies. The rate of the entropy change ($\dot{S}$) and the power ($P$) of the Hawking emission are calculated from the relevant components of the anomalous stress-tensor under the Unruh vacuum condition. I show that the dependence of $\dot{S}$ on power is $\dot{S}\propto P^{1/2}$ which is identical to that for the information flow in one dimensional system. This is established by using the ($1+1$) dimensional gravitational anomalies first. Then the fact is further bolstered by considering the ($1+3$) dimensional gravitational anomalies. It is found that in the former case, the proportionality constant is exactly identical to one dimensional situation, known as Pendry's formula, while in later situation its value decreases.     
\end{abstract}
\vskip 9mm

     The fact that the black hole has one dimensional nature as long as entropy flow is concerned, has already been broad casted by Bekenstein and Mayo \cite{Bekenstein:2001tj}. This is a complementary statement to the ``holographic principle'' \cite{Susskind:1994vu}. The idea was to calculate the rate of entropy change for the emission spectrum from the horizon. In this approach the entropy current was defined by the entropy density multiplied by the group velocity of the emitted particle. It turns out that the rate of entropy flow is proportional to the square root of the energy current or power of the emitted spectrum. The same nature also occurs for an one dimensional system, except the proportionality constant is different. In this analysis the spacetime was taken to be the ($1+3$)-dimensional Schwarzschild black hole. Recently, the analysis has been extended to arbitrary $D$-dimensional Schwarzschild spacetime \cite{Hod} and the conclusion is same except the proportionality constant now depends on dimension of the spacetime.

   In this letter, the one dimensional behavior of the black hole will be explored in the context of gravitational anomalies. The roles of anomalies have already been explored in different contexts of gravitational physics. It has been already observed that the gravitational anomalies can explain Hawking effect \cite{Christensen:1977jc}--\cite{Banerjee:2008sn}. The correct expression for the Hawking flux was derived from the anomalous stress-tensor with the suitable choice of boundary condition (see, for details \cite{Banerjee:2008wq}). Recently, the modifications to the constitutive relations for a fluid were found in presence of anomalies \cite{Son:2009tf,Majhi:2014eta,Majhi:2014hsa}. But so far, no one discussed any connection between the gravitational anomalies and one dimensional behavior of the black holes. Here I shall precisely address this issue. I shall show that the existence of anomalies at the quantum level can explain this precise property of the emission from the horizon. 

    The organization of the letter is as follows. We shall begin our discussion with the near horizon effective theory which is ($1+1$)-dimensional and hence the theory is accompanied by the two dimensional anomalies. Solutions of the anomalous equations will give the different components of stress-tensor. Using them in the Gibbs-Duhem relation \cite{Reif}, the rate of the entropy change of the emitted particle will be calculated. We will see that it is, like the one dimensional radiation, proportional to the square root of the power of the emission spectrum. Finally, this fact will be bolstered by extending the same approach for the ($1+3$)-dimensional theory in presence of gravitational anomalies.  

\vskip 3mm
   We start our analysis with the near horizon effective ($1+1$) dimensional theory where the two dimensional anomalies appear. The whole discussion will be confined within the metrics which are asymptotically flat at infinity. The reason for the restriction will be cleared later.
   Now the question is why we are interested on the two dimensional case, although the full spacetime is not ($1+1$) dimensional? The reason is as follows. 
   One of the interesting facts of black hole spacetimes is that they are effectively ($1+1$) dimensional near the event horizon \cite{Robinson:2005pd}. This can be viewed in the following way. For simplicity, consider a scalar field action with mass and interaction terms under a full black hole spacetime. The action, when expanded and then the near horizon limit has been taken, reduces to a free action for a collection of scalar field modes with the ($1+1$) dimensional background of form:
\begin{equation}
ds^2 = -f(r)dt^2 + \frac{dr^2}{f(r)}~,
\label{1.01} 
\end{equation}
where the horizon $r=r_0$ is determined by the equation $f(r_0)=0$.
Therefore, it turns out that the theory, near the horizon, is effectively a two dimensional conformal theory, because the mass term or any other interaction term vanishes. Moreover, the effective metric in this region is taken to be that given by Eq. (\ref{1.01}). Hence, here the theory is dominated by this metric (For more details, see \cite{Majhi:2011yi}). It appeared that such a fact is very useful to study several features of black holes. For instance, the Hawking effect has been discussed by two interesting approaches: one is by gravitational anomaly method \cite{Robinson:2005pd}--\cite{Banerjee:2008sn} and other is by tunnelling approach \cite{Srinivasan:1998ty}--\cite{Banerjee:2009wb}. In the first approach, it has been argued that since the near horizon theory is two dimensional, the theory must accompanied by the gravitational anomalies (two dimensional). These are manifested through the non-conservation of the energy-momentum tensor and non-vanishing of the trace of it. Then the ($t,r$) component of the stress-tensor, obtained by solving the anomaly equations under the metric (\ref{1.01}), with the Unruh boundary condition leads to the correct expression for the emission flux from the horizon. In the other approach, one considers that the tunnelling of particle is happening along radial path. So no angular part will contribute and the metric can be considered of the form given by (\ref{1.01}). Finally the correct Hawking temperature is identified by calculating the tunneling probability \cite{Srinivasan:1998ty}--\cite{Banerjee:2008cf} or the emission spectrum \cite{Banerjee:2009wb}. Also, the entropy has been calculated using the tunnelling modes \cite{Ghosh:2010pb}. So from the above evidences it is clear that the effective metric plays a central role in the black hole thermodynamics. Furthermore, the Hawking temperature has been obtained by the global embedding approach of the effective metric of the form (\ref{1.01}) in stead of the full metric \cite{Banerjee:2010ma}. Hence, we feel that it might be one of the hints in favor of the one dimensional (means one space dimension) nature of the black hole emission.

   From the above discussion, one might think that the ($1+1$) dimensional gravitation anomalies can shed some light towards the one dimensional entropy flow nature. I shall show in the following that this is indeed possible. The two dimensional (means one time and one space coordinates) gravitational anomalies are manifested by the non-conservation and non-zero value of trace of the energy-momentum tensor \cite{Bardeen:1984pm}--\cite{AlvarezGaume:1983ig}:   
\begin{equation}
\nabla_b T^{ab}=c_g\bar{\epsilon}^{ab}\nabla_b R; \,\,\,\ T^a_a = c_wR~;
\label{1.02}
\end{equation}
where $c_g$ and $C_w$ are anomaly coefficients and $R$ is the two dimensional Ricci scalar. The expressions for the components of the energy-momentum tensor, under the background (\ref{1.01}), come out to be \cite{Majhi:2014eta}
\begin{eqnarray}
&&T_{uu}=\frac{2c_g+c_w}{4}\Big(ff''-\frac{f'^2}{2}\Big)+C_{uu}~;
\nonumber
\\
&&T_{vv}=-\frac{2c_g-c_w}{4}\Big(ff''-\frac{f'^2}{2}\Big)+C_{vv}~;
\nonumber
\\
&&T_{uv} = -\frac{c_w}{4}fR=\frac{c_w}{4}ff''~;
\label{1.03}
\end{eqnarray}
where $C_{uu}$ and $C_{vv}$ are integration constants. The above are expressed in null coordinates: $u=t-r_*$ and $v=t+r_*$ with $dr_*=dr/f(r)$. Now since we are interested to the emitted particles from the horizon; i.e. the Hawking radiated particle, one must impose the Unruh Vacuum \cite{Banerjee:2008wq} to fix the integration constants. This is defined by the conditions: $T_{uu}=0$ near to the horizon $r\rightarrow r_0$ while $T_{vv}=0$ at asymptotic infinity $r\rightarrow \infty$. Imposition of the boundary conditions leads to
\begin{equation}
C_{uu} = (2c_g+c_w)2\pi^2T_0^2; \,\,\,\ C_{vv}=0~;
\label{1.04}
\end{equation}
where $T_0 = f'(r_0)/4\pi$ is the Hawking temperature. Next we shall define the energy density and entropy density by the above components.

  Now it is known that the time-time component of the stress-tensor; i.e. $T^t_t$, is related to the energy density ($\epsilon$) while the radial-radial component; i.e. $T^r_r$, is related to the pressure ($p$) of the emitted radiation. So to determine these quantities one has to express (\ref{1.03}) in original ($t,r$) coordinates. In these coordinates the components of $T_{ab}$ reduce to the following forms:
\begin{eqnarray}
&&T^t_t = -\frac{1}{f}\Big(T_{uu}+T_{vv}+2T_{uv}\Big)=-\frac{c_w}{4}\Big(4f''-\frac{f'^2}{f}\Big)-\frac{C_{uu}+C_{vv}}{f}~;
\nonumber
\\
&&T^r_r = \frac{1}{f}\Big(T_{uu}+T_{vv}-2T_{uv}\Big)=-\frac{c_w}{4}\frac{f'^2}{f}+\frac{C_{uu}+C_{vv}}{f}~;
\label{1.05}
\end{eqnarray}
If the energy density and the entropy density are denoted by $\epsilon$ and $s$, respectively, then the energy current is given by $\epsilon^a = \epsilon u^a$ while the entropy current turns out to be $s^a=su^a$ where $u^a$ is the velocity of the emitted radiation. Now if we want to calculate these quantities in the co-moving frame, then the components of the velocity for the metric (\ref{1.01}) are $u^a=(1/\sqrt{f},0)$. Hence only the time component of both the currents are non-zero. 
Now the rate of change of energy; i.e. the power ($P$), near the horizon, as measured from infinity, is 
\begin{equation}
P = \epsilon^t|_{r\rightarrow\infty} = -T^t_t u^t|_{r\rightarrow\infty} = C_{uu}+C_{vv}= (2c_g+c_w)2\pi^2T_0^2~.
\label{1.06}
\end{equation}
To calculate the rate of change of entropy, we need to find the entropy density first. Using Gibbs'-Duhem relation $T_0s = \epsilon+p$ one finds the entropy density as 
\begin{equation}
s=\frac{\epsilon+p}{T_0} = \frac{-T^t_t+T^r_r}{T_0}~.
\label{1.07}
\end{equation}
Therefore, similar to power, the rate of change of entropy, as measured from infinity, turns out to be
\begin{equation}
\dot{S}=su^t|_{r\rightarrow\infty}=\frac{2(C_{uu}+C_{vv})}{T_0} = (2c_g+c_w)4\pi^2T_0~.
\label{1.08}
\end{equation}
Note that to calculate the total rate of change of any quantity we need to integrate the corresponding density by the space volume for the unit time. In the present analysis we first calculate them near the horizon where the space dimension is one and so the rate of change of them are coming out to be the density multiplied by the velocity; i.e. the current. Next they have to be measured from infinity where $f'$ and $f''$ vanish as for asymptotic flat spacetime the form of the metric coefficient is of the form: $f(r)=1+a_0/r+\mathcal{O}(1/r^2)$. This has been precisely done in the above calculations. 
Next, combining (\ref{1.06}) and (\ref{1.08}) we find the relation between $\dot{S}$ and $P$ as
\begin{equation}
\dot{S}=\frac{2P}{T_0} = \Big[8\pi^2(2c_g+c_w)P\Big]^{1/2}~.
\label{1.09}
\end{equation}
The first equality is exactly same to what was obtained earlier for a one dimensional thermal radiation of photon (see Eq. (4) of \cite{Bekenstein:2001tj}). The second equality exhibits that $\dot S$ is proportional to square root of the power. This nature is again identical to one dimensional case. The proportionality constant, of course, depends on the value of the anomaly coefficients. It has been observed earlier that the $T^r_t$ component leads to the correct value of the Hawking flux for either $c_w=1/24\pi$ and $c_g=0$ (only trace anomaly exists) \cite{Christensen:1977jc} or $c_w=1/48\pi$ and $c_g=1/96\pi$ (both trace and diffeomorphism anomalies exist) \cite{Robinson:2005pd}--\cite{Banerjee:2008sn}. In any of the two cases we have $2c_g+c_w=1/24\pi$. Interestingly, then the proportionality constant turns out to be $(\pi/3)^{1/2}$ which is identical to that for the one dimensional radiation \cite{Bekenstein:2001tj} and hence the expression (\ref{1.09}) is exactly identical to Pendry's one \cite{Pendry}.  The reason is the following. Remember that the above analysis is valid in the near horizon limit and in this region, the effective potential of the black hole spacetime is very small. Therefore the effect of it can be neglected and hence the radiation is pure thermal. This has exactly been accounted in the above analysis and one can justify this by finding the $T^r_t$ component at infinity (which is the Hawking flux) from (\ref{1.03}). Since Pendry's work is also based on the thermal radiation, it is not surprising that the present case exactly matches with that of Pendry.  Of course, if one considers the grey body factor in the emission spectrum from the horizon then these two analysis must not give identical results.

   So far the one dimensional nature of black holes has been discussed based on the near horizon effective theory. Now in the next, I shall use the ($1+3$) dimensional gravitational anomaly and show that the radiation from the horizon has the similar nature. In literature, the expressions for components of renormalized stress-tensor corresponding to trace anomaly in the case of Schwarzschild black hole exists. Without introducing much details, let me just borrow the expressions. The stress-tensor in Unruh vacuum, in the asymptotic infinity limit $r\rightarrow\infty$, is given by \cite{Candelas:1980zt,Balbinot:1999vg},
\begin{equation}
\begin{array}{lc}
T^a_b = \frac{L}{4\pi r^2}& \left(\begin{array}{@{}cccc@{}}
                    -1 & -1 & 0 & 0 \\
                     1 & 1 & 0 & 0 \\
                    0 & 0 & 0 & 0 \\
                    0 & 0 & 0 & 0
                  \end{array}\right) \\[15pt]
\end{array},
\label{1.10}
\end{equation}
where $L$ is known as Luminosity and its value is $L=C_0/\pi M^2$, $M$ is the mass of the Schwarzschild black hole. The value of the constant $C_0$, obtained by two different methods, is \cite{DeWitt:1975ys,Elster:1983pk}
\begin{equation}
    C_0= 
\begin{cases}
    2.197\times 10^{-4},& \text{by geometric optics}\\
    2.337\times 10^{-4},& \text{by numerical estimation}~.
\end{cases}
\label{1.11}
\end{equation}
So in this case, the energy density and pressure are given by $\epsilon=-T^t_t=L/4\pi r^2=16C_0T_0^2/r^2$ and $p=T^r_r=L/4\pi r^2=16C_0T_0^2/r^2$, respectively, where $T_0 = 1/8\pi M$ is the Hawking temperature. Therefore, the entropy density is $s=(\epsilon+p)/T_0 = 32C_0T_0/r^2$. As earlier, to find the rate of change of these quantities we need to integrate them by the space volume per unit time. This is in co-moving frame is given by $4\pi r^2 u^a$. Now since $u^a=(1/\sqrt{f},0,0,0)$ the value of the unit time space volume at infinity is $4\pi r^2$. Hence the power and the rate of change of entropy of the emitted spectrum are
\begin{eqnarray}
&&P= 4\pi r^2 \epsilon = 64\pi C_0 T_0^2~;
\nonumber
\\
&&\dot{S} = 4\pi r^2 s = 128\pi C_0 T_0~.
\label{1.12}
\end{eqnarray}
Combining these two we obtain an identical relation to that of the earlier effective two dimensional case: $\dot{S} = 2P/T_0$ (see the first equality of Eq. (\ref{1.09})). Now using the expression for $T_0$ in terms of $P$ from the first relation of (\ref{1.12}) we find
\begin{equation}
\dot{S} = 16 C_0^{1/2}\Big(\pi P\Big)^{1/2}~.
\label{1.13}
\end{equation}
Here also we are getting the similar dependence of rate of change of entropy flow on the power like the one dimensional one. The only difference is the value of the proportionality constant. In the one dimensional case its value is $(\pi/3)^{1/2}$ whereas in ($1+3$) dimensions the present analysis shows that the value is
\begin{equation}
16 C_0^{1/2}\pi^{1/2} = 
\begin{cases}
    0.237\times\pi^{1/2},& \text{by geometric optics}\\
    0.245\times \pi^{1/2},& \text{by numerical estimation}~.
\end{cases}
\label{1.14}
\end{equation}
So (\ref{1.13}) is $41\%$--$42\%$ of the one dimensional situation.

   Before finishing the discussion, let me point out that Bekenstein and Mayo \cite{Bekenstein:2001tj} got a different proportionality constant in the case of four dimensional Schwarzschild black hole. Their value is $0.087\times\pi^{1/2}$. So in my analysis the rate of change of entropy increases by $2.72$--$2.82$ times compared to that obtained in \cite{Bekenstein:2001tj}. This discrepancy is due to the following fact. In ($1+3$) dimensions, the anomaly equations can not be solved exactly and hence the value of the constant $C_0$ is not exact. Whereas the value determined by Bekenstein and Mayo based on the numerical estimations done by Page \cite{Page:1976df}. In both cases different approximation techniques have been adopted and hence one can not expect identical result. 

\vskip 3mm
  Gravitational anomalies have big role in the thermodynamics of black holes \cite{Christensen:1977jc}--\cite{Banerjee:2008sn,Majhi:2012st}. Here I showed that they can revel the one dimensional nature of entropy flow rate through Hawking radiation. This has been established using both the two dimensional and four dimensional anomaly expressions. First the ($1+1$) dimensional case has been analysed by the fact that near the horizon any black hole space time is effectively two dimensional; i.e. only the ($t-r$) sector is important. It was found that the result obtained is exactly matches with the one dimensional photon emission model. Finally the fact is further bolstered by using the four dimensional anomalous stress-tensor in the case of Schwarzschild metric.

   One importance of the present analysis is that the former calculation (i.e. two dimensional) is applicable to any black holes which are asymptotically flat. This is because, as I mentioned, any static black hole spacetime is effectively of the form (\ref{1.01}) near its horizon. Therefore, within this approximation the result (\ref{1.09}) is true for any spacetime and hence we can conclude that, in general, the black holes are one dimensional so far as entropy flow is concerned. This information is completely new as the existing discussion \cite{Bekenstein:2001tj,Hod} confined only to Schwarzschild metric. Of course, this an approximate analysis and hence only the proportionality constant will vary case by case when the full spacetime will be considered. This happened here also when the analysis was repeated for four dimensional case. Moreover, the analysis is completely new which once again tells that gravitational anomalies have a leading role in thermodynamics of black holes. Let me finish with the mention of an existing work \cite{Mirza:2014vsa} which also discussed similar topic for the arbitrary $D$ space dimensional Schwarzschild black hole and black hole in Lovelock gravity. It was found that the dependence of $\dot{S}$ on $P$ is not like Pendry for Lovelock case, which, as pointed out in \cite{Hod}, is due to improper choice of radiating area. Of course, this topic is still open and one needs to further investigate to achieve any conclusive statement.    
    
\vskip 9mm
\noindent
{\bf{Acknowledgements}}\\
\noindent
The research of the author is supported by a START-UP RESEARCH GRANT (No.: SG/PHY/P/BRM/01) from Indian Institute of Technology Guwahati, India.

\end{document}